\def \abc#1#2#3#4 {\reference#1, {\sl#2}, {\bf#3}, #4}
\def \blank {\lower 5pt\hbox to 0.75in{\hrulefill}}

\def \cm{~\rm{cm}}

\def \g{~\rm{g}}

\def \yr{~\rm{yr}}
\def \K{~\rm{K}}

\def \lae{\mathrel{<\kern-1.0em\lower0.9ex\hbox{$\sim$}}}
\def \gae{\mathrel{>\kern-1.0em\lower0.9ex\hbox{$\sim$}}}

\documentstyle[12pt,preprint,epsfig]{aastex}
\begin{document}

\title{A Superwind from Early Post-Red Giant Stars?}

\shorttitle{Superwind in RGB Stars}
\shortauthors{Soker et al.}

\author{Noam Soker\altaffilmark{1}, 
M\'arcio~Catelan\altaffilmark{2}\altaffilmark{,3},
Robert~T.~Rood}
\affil{University of Virginia, Department of Astronomy, P.O.~Box 3818, 
       Charlottesville, VA 22903-0818, USA}
\author{Amos Harpaz}
\affil{University of Haifa at Oranim,
       Department of Physics, Oranim, Tivon 36006, Israel.} 
\email{soker@physics.technion.ac.il, mcatelan@astro.puc.cl, rtr@virginia.edu,
phr89ah@techunix.technion.ac.il}

\altaffiltext{1}{On Sabbatical from the University of Haifa at Oranim,
       Department of Physics, Oranim, Tivon 36006, Israel.} 

\altaffiltext{2}{Hubble Fellow.} 

\altaffiltext{3}{Present address: Pontif\'\i cia Universidad Cat\'olica de Chile, 
       Departamento de Astronom\'\i a y Astrof\'\i sica, Av. Vicu\~na Mackenna 4860, 
       Santiago 22, Chile}

\begin{abstract}

We suggest that the gap observed at $\sim 20,\!000\,$K in the horizontal
branches of several Galactic globular clusters is caused by a small
amount of extra mass loss which occurs when stars start to ``peel off''
the red giant branch (RGB), {{{ i.e., when their effective temperature  
starts to increase, even though they may still be on the RGB.}}} 
 We show that the envelope structure of RGB stars {{{ which start}}} 
to peel off is similar to that of late asymptotic
giant branch stars known to have a super-wind phase. An analogous super-wind
in the RGB peel-off stars could easily lead to the observed gap in
the distribution of the hottest HB stars. 

\end{abstract}



\section{INTRODUCTION}

Gaps along the principal sequences of globular cluster (GC) color-magnitude 
diagrams (CMDs) constitute one of the most intriguing problems in the 
evolution of low-mass stars. Sandage, Katem, \& Kristian (1968) first 
reported that a ``major significant" gap was present on the red giant 
branch (RGB) of M15 (NGC~7078), and subsequently similar features were 
reported in several other GCs. Most prominent among these have been the 
gaps along the horizontal branch (HB), which is the core helium-burning 
evolutionary phase immediately following the RGB phase. The HB is 
thought to constitute a sequence in post-RGB mass (Rood 1973), where stars 
with progressively lower masses (i.e., with higher mass loss rates on the 
RGB) end up at hotter and hotter regions on the zero-age HB (ZAHB). 

While the statistical significance of the Sandage {\it et al.} (1968) 
gap on the M15 RGB has been questioned on the basis of detailed Monte Carlo 
simulations (Bahcall \& Yahil 1972), and while similar arguments 
(Catelan {\it et al.} 1998) 
have been put forward to question the statistical significance of gaps 
similar to the  
famous ``Newell (1973) gaps" along the HB, it is still possible that some 
of the detected gaps on the HB may turn out to be real 
(Ferraro {\it et al.} 1998; Piotto {\it et al.} 1999), 
thereby requiring a physical (rather than purely statistical or 
mathematical) explanation for their occurrence. 
For instance, Brown {\it et al.} (2001) have recently provided a convincing 
physical explanation for a gap located inside the ``extreme HB" (EHB) of 
NGC~2808 (Bedin {\it et al.} 2000). 

Besides the EHB gap, one cooler gap, presumably located somewhere in the  
range $T_{\rm eff, G1} \approx 10,\!000 - 12,\!000$~K, has often been 
suggested in the literature; this is Newell's (1973) ``gap~1," or the 
Ferraro {\it et al.} (1998) 
gap ``G1." Note that this gap's temperature is intriguingly 
close to the onset of radiative levitation of heavy elements in HB stars, 
which has been shown to give rise to observable consequences in the CMDs 
of GCs, particularly in the shorter wavelengths (Grundahl {\it et al.} 1999). 
Indeed, Caloi (1999) has suggested a connection between a gap at 
$(\bv)_0 = 0$ (which however is substantially cooler than G1) 
and radiative levitation. While 
interesting, it remains to be proven that a {\em real} gap exists at the 
location of G1 in {\em all} Galactic GCs with sufficiently hot HB stars, 
in the same manner that radiative levitation of metals clearly  
{\bf leads} to 
identifiable features in the near-UV CMDs of {\em all} GCs whose HBs 
contain stars hotter than $T_{\rm eff} \approx 11,\!500$~K. The compilation 
presented by Catelan {\it et al.} (1998) shows that the fraction of GCs showing 
one such gap is relatively small, which strengthens the argument that this 
gap too may turn out to be the result of a statistical fluctuation. 

However, the same argument may not apply to Newell's (1973) ``gap~2,"
which appears to be located in the same position as gap G3 in Ferraro et
al. (1998). This is a gap located at a much hotter region of the HB, near
the boundary between the blue HB (BHB) and the EHB, at $T_{\rm eff, G3}
\approx 20,\!000$~K. While the number of GCs showing extended blue tails
reaching the G3 region is small compared with the number of GCs which have
the region around G1 populated, several authors have recently argued that
all GCs with sufficiently extended blue tails do seem to show the G3 gap
(Ferraro {\it et al.} 1998; Piotto {\it et al.} 1999; Brown {\it et al.} 2001).  
Ferraro {\it et al.} argue that such a gap occurs at roughly the same 
temperature in all clusters, $T_{\rm eff, G3} \simeq 20,\!000$~K 
(cf. their Table~2).
 In contrast, Piotto {\it et al.} argue that clusters with different 
metallicity have gap G3 located at somewhat different temperatures and 
the quantity remaining constant from one case to the next is the ZAHB 
mass of the observed gaps: $0.53\,M_\odot < M < 0.54\,M_\odot$. 
{{{ The Piotto {\it et al.}
result is based on $B,~V$ photometry which is not well suited for
determining the physical parameters of such hot stars. To firmly establish
the nature of the gap extensive ultraviolet photometry of a large sample
of blue-tail GCs is needed. What is clear now is that all clusters with a
substantial HB populations hotter than $\sim 20,000$ K observed to date
show a gap, and that the envelope mass for the stars hotter than G3 is
quite small.
 In any case, our proposed scenario does not depend strongly on the
gap being for a constant total stellar mass. 
It could as well be at a constant envelope mass, and 
have some dependence on metallicity (although the core mass 
does not depend strongly on metallicity; Sweigart 1987). }}}

 Our goal here is to explore a possible physical reason for the occurrence
of gap G3. We shall investigate the requirements on the (ill constrained)
RGB mass loss rates for a gap such as the one conjectured by 
Piotto {\it et al.} to appear.  
We propose that {{{ stars which start to peel-off the RGB}}}
experience a super-wind phase, i.e., an enhanced mass loss rate phase,
during the very late RGB phase and early post-RGB phase.  {{{ Although by
peel-off stars one usually refers to stars which have the core helium
flash during the post-RGB phase, in the present paper we use the term more
loosely, also to characterize all RGB stars which start to move to the
left on the HR diagram, i.e., whose surface temperature starts to
increase, even though they may still be on the RGB by the time they
undergo the helium core flash. }}} From the inner bright shell of
planetary nebulae and obscuration of late asymptotic giant branch (AGB)
stars, such a super-wind phase is known to exist in late AGB and early
post-AGB stars.  The AGB super-wind was postulated by Renzini (1981).  A
super-wind phase seems to occur in some red supergiants which are
progenitors of Type II narrow-line supernovae (Fransson {\it et al.} 2001).
 
 We begin in \S2 by evaluating the constraints on the RGB mass loss 
rate for the production of a gap similar to G3.
 In \S3 we discuss the envelope structure of RGB stars that are 
near the ``peel-off" stage at the tip of the RGB and post-RGB, 
comparing that against similar computations for AGB stars during their 
super-wind phase. 
 A summary and discussion are provided in \S4.

\section{THE SUPERWIND POSTULATION}

 Lower metallicity stars peel-off the RGB at higher temperatures, 
smaller radii, and have lower opacities compared with higher metallicity
RGB stars.  
 In what follows we scale parameters for the ${\rm [Fe/H]}=-1.48$ and 
$Z=6\times 10^{-4}$ model of D'Cruz {\it et al.} (1996).
This model leaves the RGB with
$L\simeq 1800\,L_\odot$, $T_{\rm eff}=4000 \K$, and $R=90\,R_\odot$.
{{{ By peel-off stars we refer to all stars which have their
surface temperature increasing on the RGB, even if a core helium flash 
occurs while they are still on the RGB. 
 This is more or less the stage where the super-wind occurs in AGB stars.
From Castellani \& Castellani (1993) we find that RGB stars start to 
peel-off when their total mass is $M \simeq 0.5-0.6 \,M_\odot$. 
 Stars that suffer core helium flash during the post-RGB phase, i.e.,
bona-fide peel-off stars, have much lower masses (D'Cruz {\it et al.} 1996).}}}
 We take the duration of the super-wind, $t_{\rm sw}$, to be similar to, or 
somewhat longer (due to slower evolution of low-mass cores) than, 
the super-wind phase in AGB stars, namely 
$t_{\rm sw} \simeq t_{\rm sw}=10^4 \yr$, and scale quantities by their 
values at the termination of the RGB.
The boundary between peel-off and He-flasher stars occurs for 
$\eta \simeq 0.7$ (D'Cruz {\it et al.} 1996; Brown {\it et al.} 2001),
where $\eta$ determines the mass loss rate via the Reimers (1975a, 
1975b) formula. 
{{{ Since we are interested in stars that start to peel off, even if
not becoming peel-off stars, we scale with a somewhat lower mass loss rate 
$\eta=0.5$}}}, to get 
\begin{equation}
\dot M_{\rm R}= - 6.5 \times 10^{-8} 
\left( \frac{\eta}{0.5} \right)
\left( \frac{L}{1800\,L_\odot} \right)
\left( \frac{R}{90\,R_\odot} \right)
\left( \frac{M}{0.5\,M_\odot}\right)^{-1}
M_\odot \yr^{-1}.
\end{equation}
Even though we use Reimers' formula as a reference, it should be noted
that any of the alternative analytical mass loss formulae discussed by
Catelan (2000) yield qualitatively similar results.  To form the observed
gap at $\sim 20,\!000 \K$, we require the peel-off stars to lose extra
mass.  Using the Piotto {\it et al.} (1999) result that the gap in different GCs
occurs at roughly the stellar mass range of $0.53\,M_\odot < M
<0.54\,M_\odot$, the required enhanced mass loss rate factor $\Gamma$ is
\begin{equation}
\Gamma= 
\frac {\Delta M_{\rm sw}}{\vert \dot M_{\rm R} \vert t_{\rm sw}} = 15
\left( \frac{\Delta M_{\rm sw}}{0.01\,M_\odot} \right)
\left( \frac{\vert \dot M_{\rm R} \vert}{6.5 \times 10^{-8}\,M_\odot \yr^{-1}}
\right)^{-1}
\left( \frac{t_{\rm sw}}{10^4 \yr} \right)^{-1}, 
\end{equation} where $\Delta M_{\rm sw}$ is the extra mass that needs to
be lost during the super-wind phase.  A similar factor, or even larger, is
thought to exist in stars leaving the AGB.  {{{ As the star loses mass,
contracts, and heats up, the super-wind ceases.  The super wind may last
longer than the scaled time here, and the enhanced mass loss factor can be
smaller, e.g., $\Gamma=3$ for $t_{\rm sw}=5\times 10^4 \yr$. In any case,
after the super-wind gradually ceases, the mass loss resumes its
``normal'' value (as provided, e.g., by Reimers' formula). Just as the
``normal'' mass loss varies from star to star due to differences in some
yet unknown parameter like rotation, the super wind probably also varies
from star to star.  The net effect of the super-wind is to increase somewhat
the amount of mass lost by the star. It will not remove the entire
envelope, just as the ``regular'' wind will not.  Only stars below the gap
have experienced a super-wind.}}} The derivation above suffers from
several large uncertainties, but it does suggest that a super-wind may
explain the G3 gap.  

\section{ENVELOPE STRUCTURE}

One of the reasons for the occurence of a super-wind in AGB stars may be 
the shallow density and steep entropy gradients in the envelope 
(Soker \& Harpaz 1999; hereafter SH99).  
To show from simple arguments that shallow density envelope profiles exist,
accompanied by a steep entropy gradient, in peel-off stars, hence we expect 
them to go through a super-wind phase, we follow the arguments of SH99. 
The photospheric pressure $P_{\rm p}$ and density $\rho_{\rm p}$ are determined
by the stellar effective temperature $T_{\rm p}$, luminosity, and photospheric
opacity $\kappa$ (e.g.,  Kippenhahn \& Weigert 1990, \S10.2).
By using the definition of the photosphere as the place where 
$\kappa \, l \rho_{\rm p}=2/3$, where $l$ is the density scale height, 
we can write 
\begin{equation}
\rho_{\rm p}= 
\frac {2}{3} 
\frac {G M \mu m_{\rm H}}{k_{\rm B}} 
\frac {1} {R^2 \kappa T_{\rm p}},
\end{equation}
where $k_{\rm B}$ is the Boltzmann constant and $\mu m_{\rm H}$ is the 
mean mass per particle.
At the level of accuracy required here, we can take the
pressure and density scale height at the photosphere to be equal.
We take the opacity from Alexander \& Ferguson (1994), for a temperature of
$T_{\rm p}\simeq 4000$~K and density of $\rho \simeq 10^{-9} \g \cm^{-3}$. 
Substituting typical values for peel-off stars, we find the ratio of
the photospheric density to the average envelope density
$\rho_{\rm a} = 3 M_{\rm e} /(4 \pi R^3)$ to be 
\begin{equation}
\frac {\rho_{\rm p}}{\rho_{\rm a}} \simeq 0.026
\left( \frac{R}{90\,R_\odot} \right)
\left( \frac{\kappa}{10^{-3} \cm^2 \g^{-1}} \right)^{-1}
\left( \frac{T_{\rm p}}{4000\,\K} \right)^{-1}
\left( \frac{M}{0.5\,M_\odot} \right)
\left( \frac{M_{\rm e}}{0.04\,M_\odot} \right)^{-1}.
\end{equation}
 Since the density increases inward, the density ratio between the
photosphere and outer envelope regions is higher (closer to unity),
hence a very shallow density gradient exists in the outer envelope
regions of upper RGB and early post-RGB stars.  In AGB stars with
parameters similar to stars with observed super-winds, SH99 find the
ratio in equation (4) to be $\sim 0.25$.  We derive a similar ratio
of $0.25$ for red supergiants which are thought to have a super-wind
(Fransson {\it et al.} 2001), e.g., VY Canis Majoris (Kastner \& Weintraub
1998, and references therein).  The structure and properties
of RGB stars with low mass envelopes are similar to those of AGB and
red supergiant stars. Although such RGB peel-off stars have steeper
gradients than the AGB and red supergiants the gradients are much
shallower than in ``normal'' stars, so it is still plausible that
these stars have a super-wind phase. 

To emphasize the similarity between RGB and AGB stars, we constructed
upper RGB and early post-RGB models with the same numerical code that
was used to build the AGB models of SH99 (where all details of the 
calculations can be found), and the same composition was used (solar).
 We find the same behavior for low metallicity stars, but for a quantitative
comparison with AGB stars we present in this short paper
the solar metallicity results.
In Figure 1 we present the relevant variables in the same form 
as in SH99, so a direct comparison can be made. 
 The upper panel is for an RGB stellar model with an envelope mass of 
$M_{\rm e}=0.1\, M_\odot$. 
 The star at this stage still climbs the RGB (it is a solar metallicity
star), and the core has not reached its maximum mass yet.  
 The middle panel presents a star with the same radius, but a 
lower envelope mass and a more developed core; 
the star is at its very early post-RGB phase. 
 The envelope in the middle panel has a shallower density and a much
steeper entropy gradient compared with the upper pannel.
The differences in the entropy gradients are largest at
the outer $\sim 20 \%$ of the envelope. 
 The lower panel presents a model with a lower envelope mass and later
in its post-RGB phase. 
 The radius and envelope mass are lower than those for the middle panel
model in such a way that the density gradients are comparable,
but the entropy gradient is much steeper in the lower panel. 
 The steepening of the entropy gradients implies a stronger convection. 
SH99 discuss plausible ways by which these changes can 
enhance the mass loss rate. 

\section{SUMMARY}
The hottest ($>20,\!000\,$K) part of the HB in Galactic globular
clusters may be populated by stars which have peeled off,
{{{ or started to peel off}}} the RGB before the helium flash, 
{{{ i.e., whose surface temperatures increased just before the 
helium core flash.}}}
We suggest that such stars may undergo a super-wind phase analogous to that 
observed in AGB stars. 
These stars have the shallow density gradient and steep entropy gradient
thought to be important in driving AGB super-winds. Scaling laws
suggest that the peel-off star super-winds could provide enough
additional mass loss to separate their helium burning progeny from the
distribution of normal HB stars whose ancestors underwent the helium flash
at the RGB tip. 

If this hypothesis is correct, all HBs with a significant population
beyond $20,\!000\,$K should have a gap in the HB distribution. The
presence of the gap should not depend on factors like cluster density or
binary fraction. Further ultraviolet photometry of a larger sample of GCs
with blue tails is necessary to determine whether the gaps are universal
and the physical parameters of the gap. Does the gap occur at constant
$T_{\rm eff}$, constant total mass, or constant envelope mass? Coupled
with a grid of theoretical models this could point to the key envelope
structures which lead to the onset and termination of a RGB super-wind.

\acknowledgements
{{{ We thank the referee for useful comments.}}} 
This research was supported in part by grants from the 
US-Israel Binational Science Foundation (N.S.) and the Celerity Foundation.
 Support for M.C. was provided 
  by NASA through Hubble Fellowship grant HF--01105.01--98A awarded by the 
  Space Telescope Science Institute (STScI), which is operated by the Association
  of Universities for Research in Astronomy, Inc., for NASA under
  contract NAS~5--26555.  RTR is partially supported by NASA Long Term
Astrophysics Grant
NAG 5-6403 and STScI grant GO-8709.

\bigskip

{\bf FIGURE CAPTIONS}

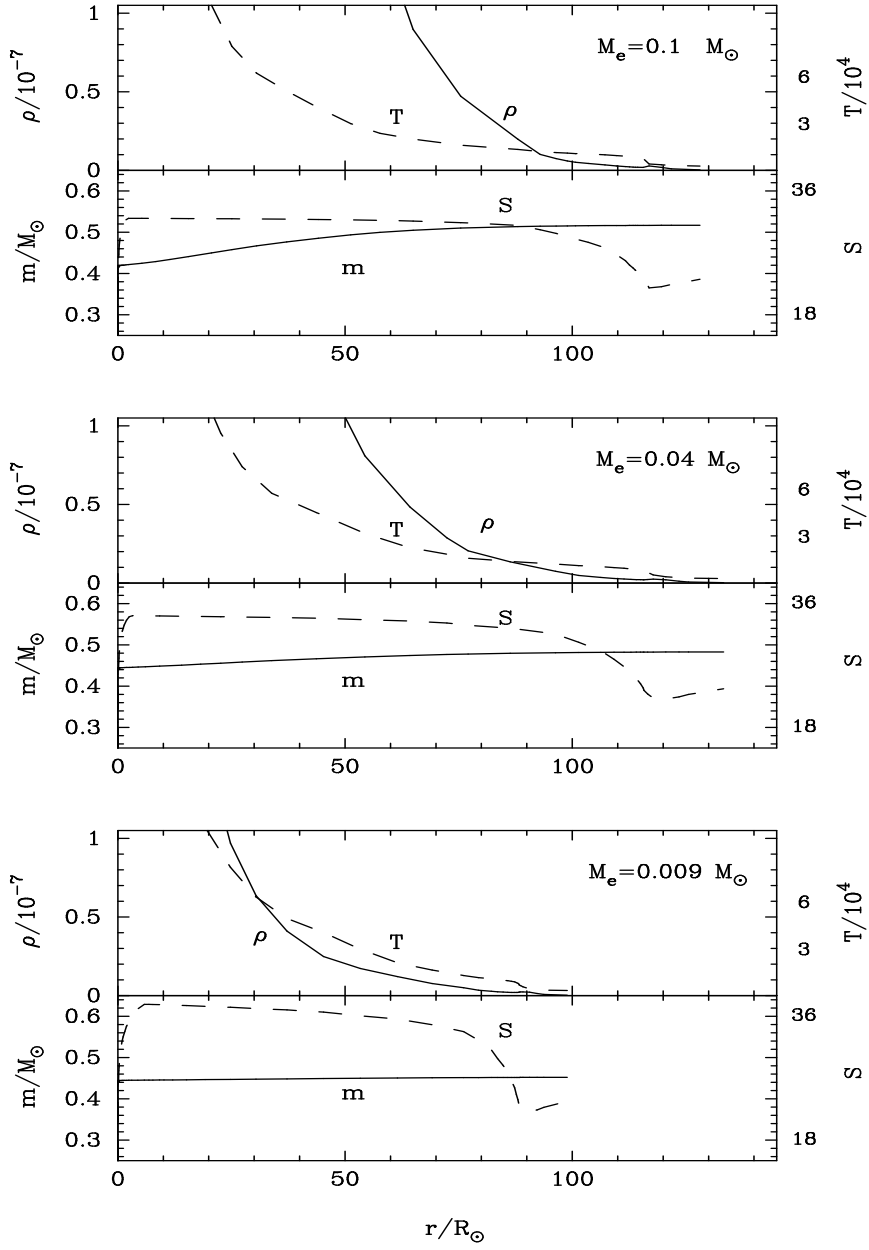
\noindent {\bf Figure 1:}
The envelope structure of three RGB and post-RGB stellar models.
The envelope mass of each model is indicated inside the panel. 
The quantities that are plotted versus the radius are: 
temperature $T$ in Kelvin, density $\rho$ in $\g \cm^{-3}$, 
the total mass $M$ in $M_\odot$, and the entropy $S$ in relative units. 
 Note that we treat the region near the photosphere using the 
Eddington approximation of gray atmosphere, and therefore 
the values of the density, pressure, and temperature 
very close to the surface (photosphere) are not accurate.
Also, these variables are drawn above the photosphere, since the numerical
code has few shells there.

\newpage

\parbox{3in}{\epsfxsize=4.5in \epsfysize=6.5in \epsfbox{sokerfig1.eps}}
\vskip 0.1in
\centerline{\parbox{4.5in}{\footnotesize  {\sc Fig.~1.---}
}
\label{fig1}
       }
\end{document}